\documentclass[graybox]{svmult}

\usepackage{mathptmx}       % selects Times Roman as basic font
\usepackage{helvet}         % selects Helvetica as sans-serif font
\usepackage{courier}        % selects Courier as typewriter font
\usepackage{type1cm}        % activate if the above 3 fonts are
\usepackage{makeidx}         % allows index generation
\usepackage{graphicx}        % standard LaTeX graphics tool
\usepackage{multicol}        % used for the two-column index
\usepackage[bottom]{footmisc}% places footnotes at page bottom

\usepackage{url}
\usepackage{acronym}
\usepackage{amssymb}
\usepackage{mathrsfs} % for Lagrangian symbol
\usepackage{newalg}
\usepackage{txfonts} % for varprod (Cartesian product)
\usepackage{subfigure}
\usepackage{color}

\makeindex             % used for the subject index
\begin{document}

\title*{Counter-Factual Reinforcement Learning: How To Model Decision-Makers That Anticipate The Future}
\titlerunning{Counter-Factual Reinforcement Learning} 
\author{Ritchie Lee, David H. Wolpert, James Bono, Scott Backhaus, Russell Bent, and Brendan Tracey}
\authorrunning{Lee, R. et al.} 
\institute{Ritchie Lee \at Carnegie Mellon University Silicon Valley, NASA Ames Research Park, Mail Stop 23-11, Moffett Field, CA, 94035 \email{ritchie.lee@sv.cmu.edu}
\and David H. Wolpert \at Santa Fe Institute, 1399 Hyde Park Rd., Santa Fe, NM 87501\\ Los Alamos National Laboratory, MS B256, Los Alamos, NM, 87545\\
\email{david.h.wolpert@gmail.com}
\and James Bono \at American University, 4400 Massachusetts Ave. NW, Washington DC 20016\\
\email{bono@american.edu}
\and Scott Backhaus \at Los Alamos National Laboratory, MS K764, Los Alamos, NM 87545\\
\email{backhaus@lanl.gov}
\and Russell Bent \at Los Alamos National Laboratory, MS C933, Los Alamos, NM 87545
\email{rbent@lanl.gov}
\and Brendan Tracey \at Stanford University, 496 Lomita Mall, Stanford, CA 94305
\email{btracey@stanford.edu}}
\maketitle

%%%%%%%%%%%%%%%%%%%%%%%%%%%%%%%%%%%%%%%%%%%%%%%%%%%%%%%%%%%%%%%%%%%%%%%%%%%%%%%%%

%\newcommand{\E}{{\mathbb{E}}}
\newcommand{\B}{{\mathscr{B}}}
\newcommand{\G}{{\mathscr{G}}}
\newcommand{\R}{{\mathbb{R}}}
\newcommand{\C}{{\mathscr{C}}}
\newcommand{\A}{{\mathscr{A}}}
\newcommand{\DD}{{\mathscr{D}}}
\newcommand{\CC}{{\mathscr{C}}}
\newcommand{\F}{{\mathscr{F}}}
\newcommand{\PP}{{\mathscr{P}}}
\newcommand{\MM}{{\mathscr{M}}}
\newcommand{\EE}{{\mathscr{E}}}
\newcommand{\btc}{\textbf}

\newacro{cpd}[CPD]{conditional probability distribution}
%\newacro{qre}[QRE]{quantal response equilibrium}
%\newacro{ne}[NE]{Nash equilibrium}
%\newacro{ch}[CH]{cognitive hierarchy}
\newacro{rl}[RL]{reinforcement learning}

%%%%%%%%%%%%%%%%%%%%%%%%%%%%%%%%%%%%%%%%%%%%%%%%%%%%%%%%%%%%%%%%%%%%%%%%%%%%%%%%%
%%%%%%%%%%%%%%%%%%%%%%%%%%%%%%%%%%%%%%%%%%%%%%%%%%%%%%%%%%%%%%%%%%%%%%%%%%%%%%%%%%

\abstract{
This paper introduces a novel framework for modeling interacting humans in a multi-stage game.  This ``iterated semi network-form game" framework
has the following desirable characteristics: (1) Bounded rational players, (2) strategic players (i.e., players account for one another's reward functions
when predicting one another's behavior), and (3) computational tractability even on real-world systems. We achieve these benefits by combining concepts from game theory and reinforcement learning.  To be precise, we extend the bounded rational ``level-K reasoning" model to apply to games over multiple stages.  Our extension allows the decomposition of the overall modeling problem into a series of smaller ones, each of which can be solved by standard reinforcement learning algorithms.  We call this hybrid approach ``level-K reinforcement learning". We investigate these ideas in a cyber battle scenario over a smart power grid and discuss the relationship between the behavior predicted by our model and what one might expect of real human defenders and attackers.

}

\acresetall
\section{Introduction}
\label{sec:intro}
\label{sec:Intro}
We are interested in modeling something that has never been modeled before: the interaction of human players in a very complicated time-extended domain, such as a cyber attack on a power grid, when the players have little or no previous experience with that domain.  Our approach combines concepts from game theory and computer science in a novel way. In particular, we introduce the first time-extended level-K game theory model \cite{Costa-Gomes06,Nagel95,Stahl95}.  We solve this model using \ac{rl} algorithms \cite{SuttonBook} to learn each player's policy against the level $K-1$ policies of the other players. The result is a non-equilibrium model of a complex and time-extended scenario where bounded-rational players interact strategically. Our model is computationally tractable even in real-world domains.

\subsection{Overview and Related Work}

The foundation of our approach is the use of a ``semi-Bayes net" to capture the functional structure of a strategic game. A semi-Bayes net is essentially a Bayes net \cite{KollerBook} where the conditional probability distributions for nodes representing player decisions are left unspecified.  Combining a semi-Bayes net with utility functions for the players yields a ``semi network-form game" (or semi net-form game)~\cite{Lee11}, which takes the place of the extensive-form game~\cite{MyersonBook} used in conventional game theory.\footnote{The ``semi-'' modifier refers to a restricted category of models within a broader class of models called network-form games. A key difference between the semi-network form game used here and the general formulation of network-form games is that the general formulation can handle unawareness -- situation where a player does not know of the possibility of some aspect of the game \cite{Wolpert12}. Unawareness is a major stumbling block of conventional game theoretic approaches in part because it forces a disequilibrium by presenting an extreme violation of the common prior assumption \cite{Halpern07}.}  In this paper, we extend the semi net-form game framework to a repeated-time structure by defining an ``iterated semi net-form game".  The conditional probability distributions at the player decision nodes are specified by combining the iterated semi net-form game with a solution concept, e.g., the level-K \ac{rl} policies used in this paper. The result is a Bayes net model of strategic behavior.

Like all Bayes nets, our model describes the conditional dependence relationships among a set of random variables. In the context of a strategic scenario, conditional dependencies can be interpreted to describe, for example, the information available to a player while making a strategic decision. In this way, semi net-form games incorporate a notion similar to that of ``information sets'' found in extensive-form games.  However, information in semi net-form games takes on the nature of information in statistics, thereby opening it to formal analysis by any number of statistical tools \cite{Kullback59,RobertBook} as opposed to information sets which uses an informal notion.  Just as information sets are the key to capturing incomplete information in extensive-form games, conditional dependence relationships are the key to capturing incomplete information in semi net-form games.\footnote{Harsanyi's Bayesian games \cite{Harsanyi67} are a special case of extensive form games in which nature first chooses the game, and this move by nature generally belongs to different information sets for the different players. This structure converts the game of incomplete information to a game of imperfect information, i.e. the players have imperfectly observed nature's move.  In addition to the fact that Harsanyi's  used extensive form games in his work while we're using semi network-form games, our work also differs in what we are modeling.  Harsanyi focused on incomplete information, while our model incorporates incomplete information and any other uncertainty or stochasticity in the strategic setting.}  In our example of a cyber battle, the cyber defender (power grid operator) has access to the full system state, whereas the cyber attacker only has access to the part of the system that has been compromised.  Representing this in the semi net-form game diagram means the defender's decision node has the full system state as its parent, while the attacker's decision node only has a subset of the state as its parent.  As a consequence, the attacker cannot distinguish between some of the system states.  In the language of extensive-form games, we say that all states mapping to the same attacker's observation belong to the same information set.

It is important to recognize that the semi net-form game model is independent of a solution concept. Just as a researcher can apply a variety of equilibrium concepts  (Nash equilibrium, subgame perfect equilibrium, quantal response equilibrium~\cite{McKelvey95,McKelvey98}, etc.) to the same extensive-form game, so too can various solution concepts apply to the same semi net-form game. In this paper we focus on the use of level-K \ac{rl} policies, however, there is no way in which the semi net-form games model is dependent on that concept. One could, in principle, apply Nash equilibrium, subgame perfect equilibrium, quantal response equilibrium, etc. to a semi net-form game, though doing so may not result in a computationally tractable model or a good description of human behavior. 

%Note that the level-K solution concept is similar to cognitive hierarcy~\cite{Camerer04}, which is essentially a level-K model where a level $K$ player has uncertainty about which levels, 0 through $K-1$, the other players use \btc{why does this comment fit here?}.

In the remainder of this introduction, we describe three characteristics whose unique combination is the contribution of our paper.  The first is that players in our model are strategic; that their policy choices depend on the reward functions of the other players. This is in contrast to learning-in-games and co-evolution models \cite{Fudenberg98,Kandori93} wherein players do not use information about their opponents' reward function to predict their opponents' decisions and choose their own actions. On this point, we are following experimental studies \cite{CamererBook}, which routinely demonstrate the responsiveness of player behavior to changes in the rewards of other players.

Second, our approach is computationally feasible even on real-world problems. This is in contrast to equilibrium models such as subgame perfect equilibrium and quantal response equilibrium.  We avoid the computational problems associated with solving for equilibria by using the level-K \ac{rl} policy model, which is a non-equilibrium solution concept. That is, since level-K players are not forced to have correct beliefs about the actions of the other players, the level-K strategy of player $i$ does not depend on the actual strategy of $i$'s opponents. As a result, this means that the level-K \ac{rl} policies of each of the players can be solved independently. The computational tractability of our model is also in contrast to partially observable Markov decision process- (POMDP-) based models (e.g. Interactive-POMDPs~\cite{Doshi05}) in which players are required to maintain belief states over belief states thus causing a quick explosion of the computational space.  We circumvent this explosion of belief states by formulating policies as mappings from a player's memory to actions, where memory refers to some subset of a player's current and past observations, past actions, and statistics derived from those variables.  This formulation puts our work more squarely in the literature of standard \ac{rl} \cite{KaelblingSurvey,SuttonBook}.  As a final point of computational tractability, our approach uses the policy representation instead of the strategic representation of player decisions.  The difference is that the policy representation forces player behavior to be stationary -- the time index is not an argument of the policy -- whereas in the strategic representation strategies are non-stationary in general.  

Third, since our goal is to predict the behavior of real human players, we rely heavily on the experimental game theory literature to motivate our modeling choices. Using the policy mapping from memories to actions, it is straightforward to introduce experimentally motivated behavioral features such as noisy, sampled or bounded memory.  The result of the \ac{rl}, then, is an optimal strategy given more realistic assumptions about the limitations of human beings.\footnote{One can imagine an extension where the \ac{rl} training is modified to reflect bounded rationality, satisfying \cite{Simon56}, etc. For example, to capture satisficing, the \ac{rl} may be stopped upon achieving the satisficing level of utility.  Note that we do not pursue such bounded rational \ac{rl} here.}  This is in contrast to the literature on coevolutionary \ac{rl} \cite{Fogel06,Moriarty99}, where the goal is to find optimal strategies.  For example, the work in \cite{Fogel01} uses \ac{rl} to design expert checkers strategies. In those models, behavioral features motivated by human experimental data are not included due to the constraining effect they have on optimal strategies. Hence, \ac{rl} in our model is used as a description of how real humans behave. This use for \ac{rl} has a foundation in neurological research \cite{Dayan02,Maia09}, where it has provided a useful framework for studying and predicting conditioning, habits, goal-directed actions, incentive salience, motivation and vigor \cite{Maia11}.  The level-K model is itself another way in which we incorporate experimentally motivated themes. In particular, by using the level-K model instead of an equilibrium solution concept, we avoid the awkward assumption that players' predictions about each other are always correct \cite{CamererBook,Kagel95,Plott08}.

We investigate all of this for modeling a cyber battle over a smart power grid.  We discuss the relationship between the behavior predicted by our model and what one might expect of real human defenders and attackers.

\subsection{Roadmap}

This chapter is organized as follows. In Section 2, we provide a review of semi network-form games and the level-K d-relaxed strategies solution concept~\cite{Lee11}.  This review is the starting point for the theoretical advances of this paper found in Section 3.  In Section 3 we extend the semi net-form games formalism to iterated semi network-form games, which enables interactions over a time-repeated structure.  This is also where we introduce the level-K \ac{rl} solution concept.  Section 3 is the major theoretical contribution of this paper.  In Section 4, we apply the iterated semi net-form game framework to model a cyber battle on a smart power distribution network.  The goal of Section 4 is to illustrate how an iterated semi net-form game is realized and how the level-K \ac{rl} policy solution concept is implemented. In this section we describe the setting of the scenario and lay out the iterated semi net-form game model, including observations, memories, moves and utility functions for both players. We also describe the details of the level-K \ac{rl} algorithm we use to solve for players' policies. This section concludes with simulation results and a possible explanation for the resulting behaviors.  Section 5 provides a concluding discussion of the iterated semi net-form games framework and future work.

\section{Semi Network-Form Games Review}
\label{sec:snfg-review}
In this section, we provide a brief review of semi net-form games.  For a rigorous treatment, please refer to Lee and Wolpert~\cite{Lee11}.

\subsection{Framework Description}
\label{sec:descSnfg}

A ``semi network-form game" (or semi net-form game) uses a Bayes net~\cite{KollerBook} to serve as the underlying probabilistic framework, consequently representing all parts of the system using random variables.  Non-human components such as automation and physical systems are described using ``chance" nodes, while human components are described using ``decision" nodes.  Formally, chance nodes differ from decision nodes in that their conditional probability distributions are prespecified.  In contrast, each decision node is associated with a utility function which maps an instantiation of the net to a real number quantifying the player's utility.  In addition to knowing the conditional distributions at the chance nodes, we must also determine the conditional distributions at the decision nodes to fully specify the Bayes net. We will discuss how to arrive at the players' conditional distributions over possible actions, also called their ``strategies", later in Section~\ref{sec:LKRelaxed}.  The discussion is in terms of countable spaces, but much of the discussion carries over to the uncountable case.  We describe a semi net-form game as follows:  	

An {\bf{($N$-player) semi network-form game}} is described by a quintuple $(G, X, u, R, \pi)$ where
\begin{enumerate}

\item $G$ is a finite directed acyclic graph represented by a set of vertices and a set of edges.  The graph $G$ defines the topology of the Bayes network, thus specifying the random variables as well as the relationships between them.

\item $X$ is a Cartesian product of the variable space of all vertices.  Thus $X$ contains all instantiations of the Bayes network.

\item $u$ is a function that takes an instantiation of the Bayes network as input and outputs a vector in $\mathbb{R}^N$, where component $i$ of the output vector represents player $i$'s utility of the input instantiation.  We will typically view it as a set of $N$ utility functions where each one maps an instantiation of the network to a real number.

\item $R$ is a partition of the vertices into $N+1$ subsets.  The first $N$ partitions contain exactly one vertex, and are used to associate assignments of decision nodes to players.  In other words, each player controls a single decision node.  The $N+1$ partition contains the remainder of the vertices, which are the chance nodes.

\item $\pi$ is a function that assigns to every chance node a conditional probability distribution~\cite{KollerBook} of that node
conditioned on the values of its parents.
\end{enumerate}

Specifically, ${X_v}$ is the set of all possible states at node $v$, $u_i$ is the utility function of player $i$, $R(i)$ is the decision node set by player $i$, and $\pi$ is the fixed set of distributions at chance nodes.  Semi net-form game is a general framework that has broad modeling capabilities.  As an example, a normal-form game~\cite{MyersonBook} is a semi net-form game that has no edges. As another example, let $v$ be a decision node of player $i$ that has one parent, $v'$. Then the conditional distribution $P(X_v \mid X_{v'})$ is a generalization of an information set.

\subsection{Solution Concept: Level-K D-Relaxed Strategies}
\label{sec:LKRelaxed}

In order to make meaningful predictions of the outcomes of the games, we must solve for the strategies of the players by converting the utility function at each decision node into a conditional probability distribution over that node.  This is accomplished using a set of formal rules and assumptions applied to the players called a solution concept.   A number of solution concepts have been proposed in the game theory literature.  Many of which show promise in modeling real human behavior in game theory experiments, such as level-K thinking, quantal response equilibrium, and cognitive hierarchy.  Although this work uses level-K exclusively, we are by no means wedded to this equilibrium concept.  In fact, semi net-form games can be adapted to use other models, such as Nash equilibrium, quantal response equilibrium, quantal level-K, and cognitive hierarchy.  Studies \cite{CamererBook,Wright10} have found that performance of an equilibrium concept varies a fair amount depending on the game.  Thus it may be wise to use different equilibrium concepts for different problems.

Level-K thinking~\cite{Crawford07} is a game theoretic solution concept used to predict the outcome of human-human interactions.  A number of studies \cite{Camerer10,Camerer89,CamererBook, CostaGomes09, Crawford07,Wright10} have shown promising results predicting experimental data in games using this method.  The concept of level-K is defined recursively as follows. A level $K$ player plays (picks his action) as though all other players are playing at level $K-1$, who, in turn, play as though all other players are playing at level $K-2$, etc.  This process continues until level 0 is reached, where the player plays according to a prespecified prior distribution.  Notice that running this process for a player at $K \ge 2$ results in ricocheting between players.  For example, if player A is a level 2 player, he plays as though player B is a level 1 player, who in turn plays as though player A is a level 0 player.  Note that player B in this example may not be a level 1 player in reality -- only that player A assumes him to be during his reasoning process.  Since this ricocheting process between levels takes place entirely in the player's mind, no wall clock time is counted (we do not consider the time it takes for a human to run through his reasoning process).  We do not claim that humans actually think in this manner, but rather that this process serves as a good model for predicting the outcome of interactions at the aggregate level.  In most games, the player's level $K$ is a fairly low number for humans; experimental studies \cite{CamererBook} have found $K$ to be somewhere between 1 and 2.

In \cite{Lee11}, the authors propose a novel solution concept called ``level-K d-relaxed strategies" that adapts the traditional level-K concept to semi network-form games.  The algorithm proceeds as follows.  To form the best response of a decision node $v$, the associated player $i = R^{-1}(v)$ will want to calculate quantities of the form argmax$_{x_v} [\mathbb{E}(u_i \mid x_v, x_{pa(v)})]$, where $u_i$ is the player's utility, $x_v$ is the variable set by the player (i.e., his move), and $x_{pa(v)}$ is the realization of his parents that he observes.  We hypothesize that he (behaves as though he) approximates this calculation in several steps. First, he samples $M$ candidate moves from a ``satisficing" distribution (a prior distribution over his moves).  Then, for each candidate move, he estimates the expected utility resulting from playing that move by sampling $M'$ times the posterior probability distribution over the entire Bayes net given his parents and his actions (which accounts for what he knows and controls), and computing the sample expectation $\hat{u}^K_i$.  Decision nodes of other players are assumed to be playing at a fixed conditional probability distribution computed at level $K-1$.  Finally, the player picks the move that has the highest estimated expected utility.  In other words, the player performs a finite-sample inference of his utility function using the information available to him, then picks (out of a subset of all his moves) the move that yields the highest expected utility.  For better computational performance, the algorithm reuses certain sample sets by exploiting the d-separation property of Bayes nets~\cite{KollerBook}.  The solution concept was used to model pilot behavior in a mid-air encounter scenario, and showed reasonable behavioral results.

\section{Iterated Semi Network-Form Games}
\label{sec:isnfg}
In the previous section, we described a method to model a single-shot scenario.  That is, a scenario in which each player makes a single decision.  However, most real-world scenarios are not single-shot.  Rather, what is typically seen is that the outcome is determined by a series of decisions made by each player over a time-repeated structure.  One way to model time extension is to ignore the structure, create a large ``rolled-out" net\footnote{Here we are violating the definition of a semi net-form game that each player can only control a single decision node.  One way to deal with this is to treat past and future selves as different players, but having the same utility function.} that explicitly enumerates the repeated nodes, then apply level-K d-relaxed strategies described in Section~\ref{sec:LKRelaxed}.  The problem with such an approach is that the roll-out causes a linear explosion in the number of decision nodes with the number of time steps.  Since the computational complexity of level-K d-relaxed strategies is polynomial (to the $K^{th}$ power) in the number of decision nodes~\cite{Lee11}, the algorithm becomes prohibitively slow in solving scenarios with more than a few time steps.

In this section, we extend the semi network-form game from Section~\ref{sec:snfg-review} to an ``iterated semi network-form game" (or iterated semi net-form game) in order to explicitly model the repeated-time structure of the game.  Then we introduce a novel solution concept called ``level-K reinforcement learning" that adapts level-K thinking to the iterated semi network-form game setting.

\subsection{Construction of an Iterated Semi Network-Form Game}
\label{ssec:isnfg}

\begin{figure}[ht]
\begin{minipage}[b]{0.5\linewidth}
\centering
\subfigure[]{
\includegraphics[trim=0cm 5cm 0cm 0cm, clip=true, width=0.9\linewidth]{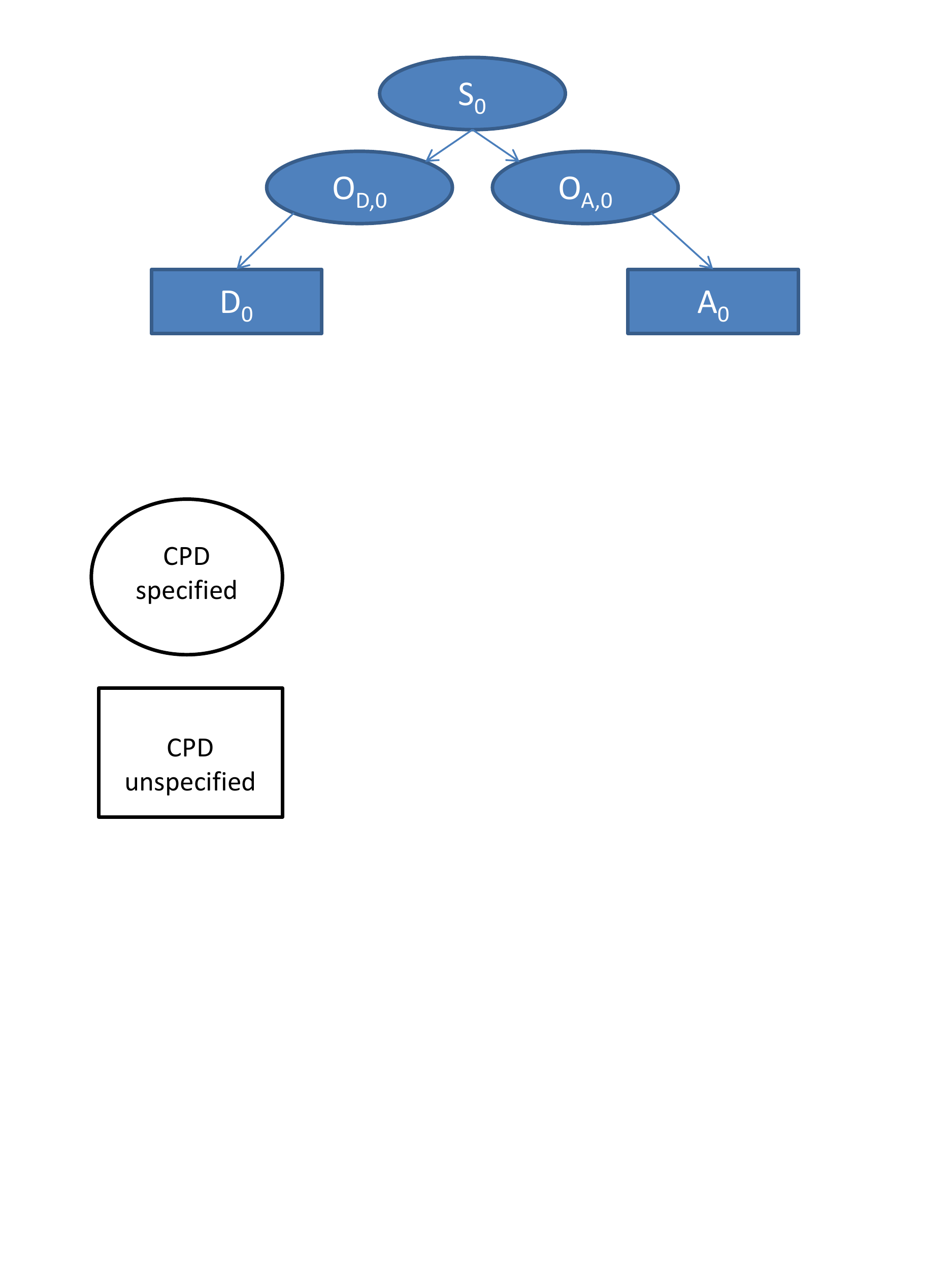}
\label{fig:sbnglue1}
}
\end{minipage}
\begin{minipage}[b]{0.5\linewidth}
\centering
\subfigure[]{
\includegraphics[trim=0cm 5cm 0cm 0cm, clip=true, width=0.9\linewidth]{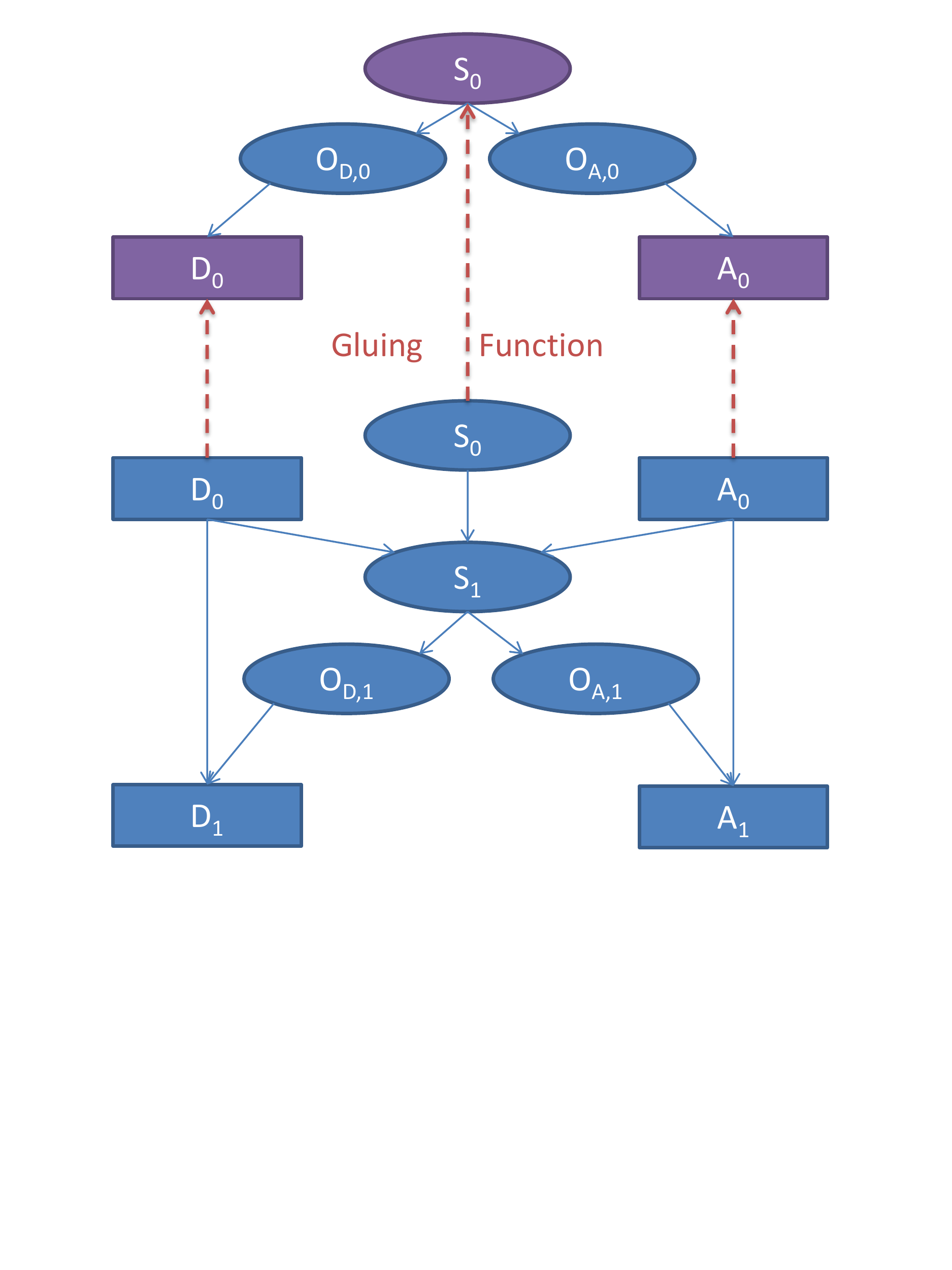}
\label{fig:sbnglue2}
}
\end{minipage}
\begin{minipage}[b]{0.5\linewidth}
\centering
\subfigure[]{
\includegraphics[width=0.9\linewidth]{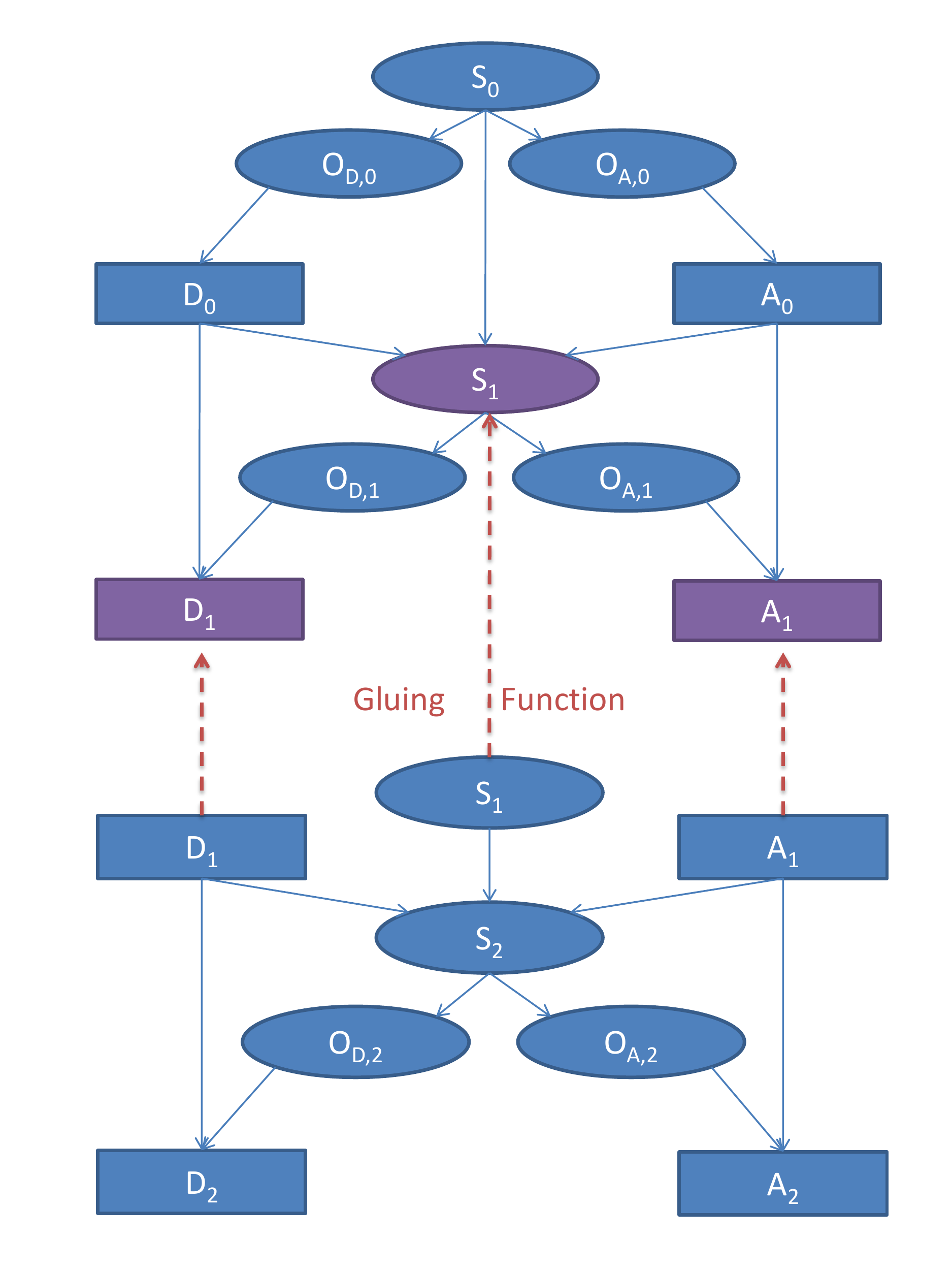}
\label{fig:sbnglue3}
}
\end{minipage}
\begin{minipage}[b]{0.5\linewidth}
\centering
\subfigure[]{
\includegraphics[width=0.9\linewidth]{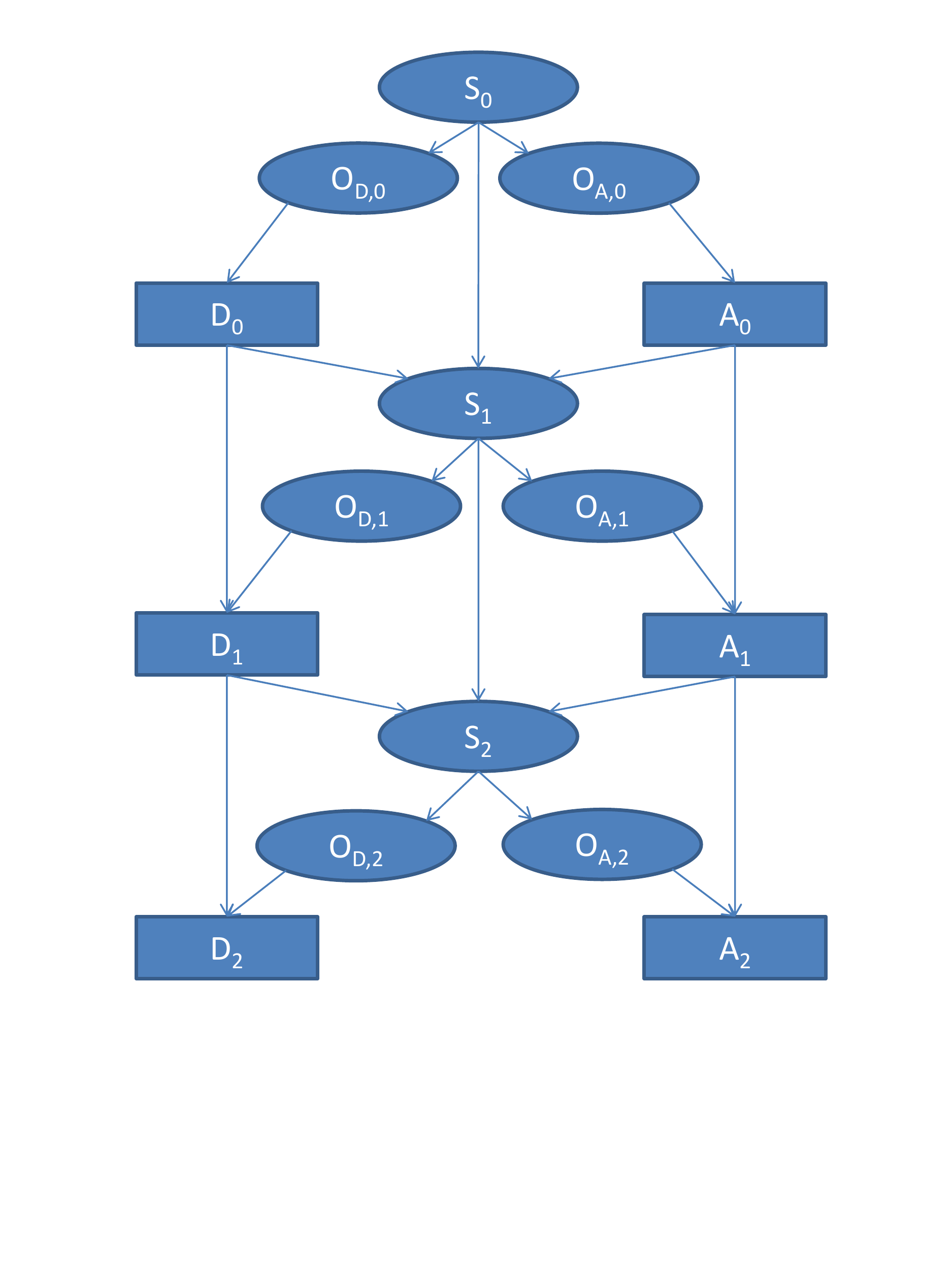}
\label{fig:sbnglue4}
}
\end{minipage}
\caption{Example construction of an iterated semi Bayes net with a base net and two kernels, i.e. $T=2$, by repeatedly applying the ``gluing" procedure.  \subref{fig:sbnglue1} A base semi Bayes net. \subref{fig:sbnglue2} A kernel semi Bayes net being ``glued" to a base semi Bayes net. \subref{fig:sbnglue3} A second kernel semi Bayes net being appended to the net. \subref{fig:sbnglue4} The final semi iterated Bayes net with $T=2$.  The numeric subscript indicates the time step to which each variable belongs.}
\label{fig:isnfgGluing}
\end{figure}

We describe the extended framework by building up the components incrementally.  A ``semi Bayes net" is like a standard Bayes net, in that a semi Bayes net has a topology specified by a set of vertices and directed edges, and variable spaces that define the possible values each vertex can take on.  However, unlike a standard Bayes net, some nodes have \acp{cpd} specified, whereas some do not.  The nodes that do not have their \acp{cpd} specified are decision nodes with one node assigned to each player.  A pictorial example of a semi Bayes net is shown in Figure \ref{fig:sbnglue1}.  The dependencies between variables are represented by directed edges.  The oval nodes are chance nodes and have their \acp{cpd} specified; the rectangular nodes are decision nodes and have their \acp{cpd} unspecified.  In this paper, the unspecified distributions will be set by the interacting players and are specified by the solution concept.

%\btc{There seems to be a lot of overlap between this section and the previous one. Is all this repetition necessary?}

We create two types of semi Bayes nets: a ``base semi Bayes net" and a ``kernel semi Bayes net".  A ``base semi Bayes net" specifies the information available to all the players at the start of play, and is where the policy decisions of the game are made.  Note that even though the game is time-extended, players only ever make one real decision. This decision concerns which policy to play, and it is made at the beginning of the game in the base semi Bayes net.  After the policy decision is made, action decisions are merely the result of evaluating the policy at the current state.  In contrast, the ``kernel semi Bayes net" specifies both how information from the past proceeds to future instances of the players during play, and how the state of nature evolves during play.  In particular, it specifies not only what a player currently observes, but also what they remember from their past observations and past actions.  For example, the kernel semi Bayes net describes how the policy chosen in the base semi Bayes net is propagated to a player's future decision nodes, where a player's action choices are merely the result of evaluating that policy. From these two, we construct an ``iterated semi Bayes net" by starting with the base semi Bayes net then repeatedly appending the kernel semi Bayes net to it $T$ times.  Each append operation uses a ``gluing" procedure that merges nodes from the first semi Bayes net to root nodes with the same spaces in the second semi Bayes net.  Figure~\ref{fig:isnfgGluing} illustrates how we build up an iterated semi Bayes net with a base net and two kernels, i.e. $T=2$.  Finally, we create an ``iterated semi net-form game" by endowing an iterated semi Bayes net with a reward function, one for each player, defined at each time instant.  The reward function takes as input an instantiation of the net at a particular (discrete) time and outputs a reward metric representing how happy the player is with that instantiation.\footnote{We use the term reward function to conform to the language used in the \ac{rl} literature. This is identical to the game theoretic notion of instantaneous utility (as opposed to the total utility, i.e. the present discounted value of instantaneous utilities).}

\subsection{Solution Concept: Level-K Reinforcement Learning}
\label{ssec:lkrl}

We introduce a novel solution concept for iterated semi net-form games that combines level-K thinking and reinforcement learning.  Instead of considering all possible combinations of actions at individual decision nodes, we simplify the decision space by assuming that the players make only a single decision -- what policy to play for the rest of the net.  That is, the players pick a policy in the base semi Bayes net, and then executes that policy over all repetitions of the kernel semi Bayes net.  This assumption allows us to convert the problem of computing a combination of actions over all time steps to one where we calculate a player's policy only once and reuse it $T$ times.  By reusing the policy, the computational complexity becomes independent of the total number of time steps.  Formally, each unspecified node of a player contains two parts: A policy and an action.  The policy is chosen in the base stage and is passed unchanged from the player's node in the base semi Bayes net to the player's node in the kernel semi Bayes net for all time steps.  At each time step, the action component of the node is sampled from the policy based on the actual values of the node's parents.  We point out that the utility of a particular policy depends on the policy decisions of other players because the reward functions of both players depend on the variables in the net. 

The manner in which players make decisions given this coupling is specified by the solution concept. In this work we handle the interaction between players by extending standard level-K thinking from action space to policy space.  That is, instead of choosing the best level $K$ action (assuming other players are choosing the best level $K-1$ action), players choose the best level $K$ policy (assuming that other players choose their best level $K-1$ policy).  Instead of prespecifying a level 0 distribution over actions, we now specify a level 0 distribution over policies.  Notice that from the perspective of a level $K$ player, the behavior of the level $K-1$ opponents is identical to a chance node.  Thus, to the player deciding his policy, the other players are just a part of his environment.  Now what remains to be done is to calculate the best response policy of the player.  In level-K reinforcement learning, we choose the utility of a player to be the sum of his rewards from each time step.  In other words, the player selects the policy which leads to the highest expected infinite sum of discounted rewards.  Noting this together with the fact that the actions of other players are identical to a stochastic environment, we see that the optimization is the same as a single-agent reinforcement learning problem where an agent must maximize his reward by observing his environment and choosing appropriate actions.  There are many standard reinforcement learning techniques that can be used to solve such a problem~\cite{BusoniuBook,KaelblingSurvey,SuttonBook}. The techniques we use in this paper are described in detail in Section \ref{sssec:rldetails}.

For example, in a two-player iterated semi network-form game, the level 1 policy of player A is trained using reinforcement learning by assuming an environment that includes a player B playing a level 0 policy. If A is instead at level 2, his environment includes player B using a level 1 policy. Player A imagines this level 1 policy as having been reinforcement learned against a level 0 player A.  To save computation time, it is assumed that how player B learns his level 1 distribution and how A imagines B to learn his level 1 distribution are identical.

\section{Application: Cyber-Physical Security of a Power Network}
\label{sec:grid}
\subsection{Introduction}
\label{ssec:gridIntro}

We test our iterated semi net-form game modeling concept on a simplified model of an electrical power grid controlled by a Supervisory Control and Data Acquisition (SCADA) system \cite{SCADA_controls_2005}.  A SCADA system forms the cyber and communication components of many critical cyber physical infrastructures, e.g., electrical power grids, chemical and nuclear plants, transportation systems, and water systems.  Human operators use SCADA systems to receive data from and send control signals to physical devices such as circuit breakers and power generators in the electrical grid. These signals cause physical changes in the infrastructure such as ramping electrical power generation levels to maintain grid stability or modifying the electrical grid's topology to maintain the grid's resilience to random component failures. If a SCADA system is compromised by a cyber attack, the human attacker may alter these control signals with the intention of degrading operations or causing permanent, widespread damage to the physical infrastructure.

The increasing connection of SCADA to other cyber systems and the use of computer systems for SCADA platforms is creating new vulnerabilities of SCADA to cyber attack \cite{Cardenas_2008}.  These vulnerabilities increase the likelihood that the SCADA systems can and will be penetrated.  However, even when a human attacker has gained some control over the physical components, the human operators still have some SCADA observation and control capability.  The operators can use this capability to anticipate and counter the attacker moves to limit or deny the damage and maintain continuity of the infrastructure's operation.  Traditional cyber security research on cyber systems has focused on identifying vulnerabilities and how to mitigate those vulnerabilities.  Here, instead, we assume that an attacker has penetrated the system, and we want to predict the outcome.

The SCADA attack and the defense by the SCADA operator can be modeled as a machine-mediated, human-human adversarial game.  In the remainder of this section, we construct an iterated semi network-form game to model just such an interaction taking place over a simplified model of a SCADA-controlled electrical grid.  The game is simulated using the level-K reinforcement learning solution concept described earlier.  We explore how the strategic thinking embodied in level-K reinforcement learning affects the player performance and outcomes between players of different level $K$.

\subsection{Scenario Model}
\label{ssec:gridEquations}

\begin{figure}[h]
\begin{center}
%\framebox[4.0in]{$\;$}
%\fbox{\rule[-.5cm]{0cm}{4cm} \rule[-.5cm]{4cm}{0cm}}
\includegraphics[width=.85\linewidth]{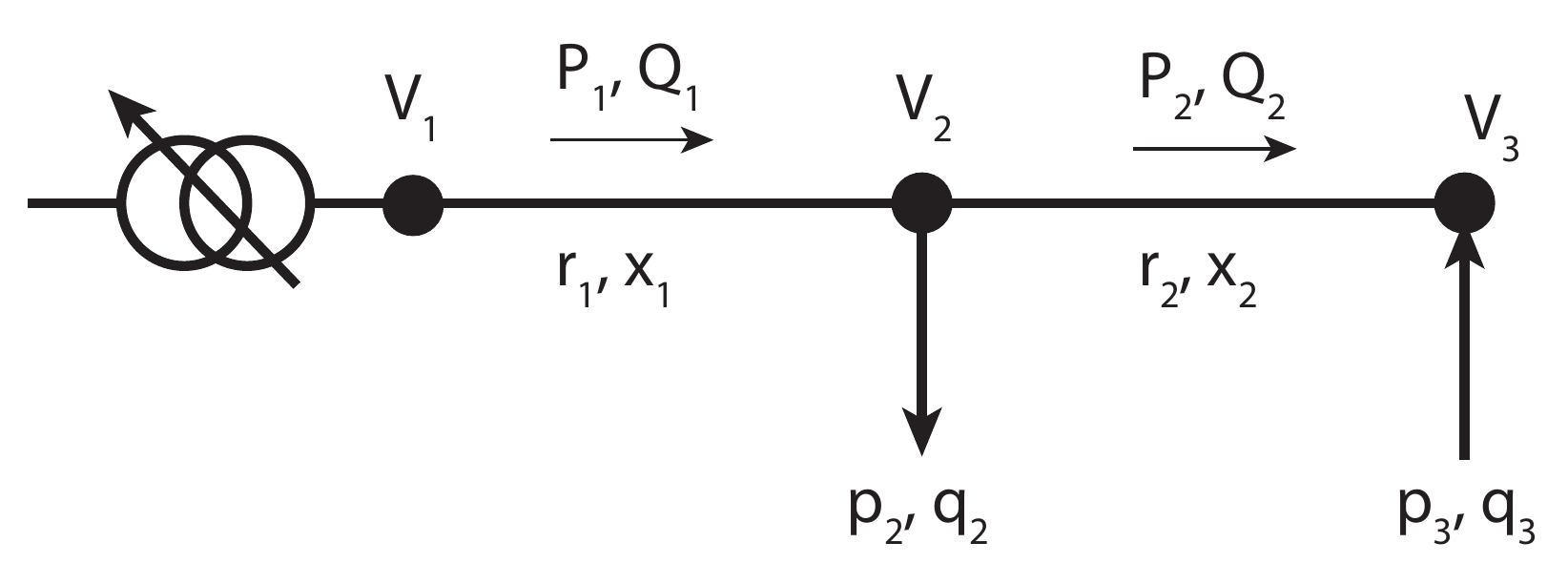}
\end{center}
\caption{Schematic drawing of the three-node distribution circuit consisting of three nodes $i$. The voltage at each node is $V_i$; the real and reactive power injections are $p_i$ and $q_i$, respectively; the line reactance and resistance are $x_i$ and $r_i$, respectively; and the real and reactive power flows in the distribution lines are $P_i$ and $Q_i$, respectively.  }
\label{fig:circuit}
\end{figure}

Figure~\ref{fig:circuit} shows a schematic of our simplified electrical grid infrastructure.  It consists of a single, radial distribution circuit \cite{PV_inverter_Q_2011} starting at the low-voltage side of a transformer at a substation (node 1) and serving customers at nodes 2 and 3.  Node 2 represents an aggregation of small consumer loads distributed along the circuit--such load aggregation is often done to reduce model complexity when simulating electrical distribution systems.  Node 3 represents a relatively large, individually-modeled distributed generator located near the end of the circuit.  

In this figure, $V_i,p_i,$ and  $q_i$ are the voltage and real and reactive power injections at node $i$.  $P_i, Q_i, r_i,$ and $x_i$ are the real power flow, reactive power flow, resistance, and reactance of circuit segment $i$.  These quasi-static power injections, power flows, voltages, and line properties are related by the nonlinear AC power flow equations~\cite{kundar}.  Our focus in this work is on the game theoretic aspects of the model, therefore, we use a linearized description of the electrical power flow, i.e., the {\it LinDistFlow} equations~\cite{PV_inverter_Q_2011}
\begin{eqnarray}
\label{eq:powerEquation1}
P_2=-p_3,\;\; Q_2=-q_3,\;\;P_1=P_2+p_2,\;\;Q_1=Q_2+q_2\\
\label{eq:powerEquation2}
V_2=V_1-(r_1 P_1 + x_1 Q_1),\;\;V_3=V_2-(r_2 P_2 + x_2 Q_2).
\end{eqnarray}
Here, all terms have been normalized by the nominal system voltage $V_0$~\cite{kundar}.

In this model, we assume that the circuit configuration is constant with $r_i=0.03$ and $x_i=0.03$. To emulate the normal fluctuations of consumer real load, $p_2$ is drawn from a uniform distribution over the range $[1.35,1.5]$ at each time step of the game.  The consumer reactive power is assumed to scale with real power, and we take $q_2=0.5 p_2$ at each step of the game. The node 3 real power injection $p_3=1$ is also taken as constant implying that, although the distributed generator at node 3 is controllable (as opposed to a fluctuating renewable generator), its output has been fixed.  Node 3 is then a reasonable model of an internal combustion engine/generator set burning diesel or perhaps methane derived from landfill gas.  Such distributed generation is becoming more common in electrical distribution systems.

In our simplified game, the SCADA operator (defender) has one objective, i.e., keeping the voltages $V_2$ and $V_3$ within appropriate operating bounds (described in more detail below).  To accomplish this the operator {\it normally} has two controls: 1) he can change the voltage $V_1$ at the head of the circuit, and 2) he can adjust the reactive power output $q_3$ of the distributed generator at node 3.  However, we assume that the system has been compromised, and the attacker has taken control of $q_3$ while the defender retains control of $V_1$.  In this circumstance, the {\it attacker }may use the injection of reactive power $q_3$ to modify all the $Q_i$ causing the voltage $V_2$ to deviate significantly from $1.0$. Excessive deviation of $V_2$ or $V_3$ can  damage customer equipment~\cite{kundar} or perhaps initiate a cascading failure beyond the circuit in question. In the language of an iterated semi network-form game, the change in $V_1$ is the decision variable of the defender, $q_3$ is the decision variable of the attacker, and $V_2$, $V_3$, and the rest of the system state are determined by the {\it LinDistFlow} equations and probability distribution described above.

\subsubsection{Players' Decision Spaces}
\label{sssec:decisionSpaces}

In this scenario, the defender maintains control of $V_1$ which he can adjust in discrete steps via a variable-tap transformer~\cite{kundar}, however, hardware-imposed limits constrain the defender's actions at time $t$ to the following domain
\begin{equation}
\label{eq:defenderDecision}
D_{D,t} = \{ \min(v_{max}, V_{1,t} + \delta v )   , V_{1,t},  \max(v_{min}, V_{1,t} - \delta v ) \}
\label{eq:defendermovespace}
\end{equation}
where $\delta v$ is the voltage step size for the transformer, and $v_{min}$ and $v_{max}$ represent the absolute min and max voltage the transformer can produce. In simple terms, the defender may leave $V_1$ unchanged or move it up or down by $\delta v$ as long as $V_1$ stays within the range $[v_{min},v_{max}]$.  In our model, we take $v_{min}=0.90$, $v_{max}=1.10$, and $\delta v=0.02$.  Similarly, hardware limitations of the generator at node 3 constrain the attacker's range of control of $q_3$.  In reality, the maximum and minimum values of $q_3$ can be a complicated function~\cite{kundar} of the maximum real power generation capability $p_{3,max}$ and the actual generation level $p_3$.  To keep the focus on the game theoretic aspects of the model, we simplify this dependence by taking the attacker's $q_3$ control domain to be
\begin{equation}
\label{eq:attackerDecision}
D_{A,t} = \{- q_{3,max}, \ldots, 0, \ldots, q_{3,max}\},
\end{equation}
with $q_{3,max}=p_{3,max}$.  To reduce the complexity of the reinforcement learning computations, we also discretize the attacker's move space to eleven equally-spaced settings with $-q_{3,max}$ and $+q_{3,max}$ as the end points.  Later, we study how the behavior and performance of the attacker depends on the size of the assets under his control by varying $p_3$ from $0$ to $1.8$.

\subsubsection{Players' Observed Spaces}
\label{ssec:observedSpaces}

 The defender and attacker make observations of the system state via the SCADA system and the attacker's compromise of node 3, respectively.  Via the SCADA system, the defender retains wide system visibility of the variables important to his operation of the system, i.e., the defender's observed space is given by
\begin{equation}
    \Omega_D=[V_1,V_2,V_3,P_1,Q_1,\mathcal{M}_D].
\end{equation}
Because he does not have access to the full SCADA system, the attacker's observed space is somewhat more limited
\begin{equation}
\Omega_A= [V_2,V_3,p_3,q_3,\mathcal{M}_A].
\end{equation}
Here, $\mathcal{M}_D$ and $\mathcal{M}_A$ each denote real numbers that represent a handcrafted summary metric of the respective player's memory of the past events in the game.  These are described in Section~\ref{sssec:memory}.

\subsubsection{Players' Rewards}
\label{sssec:rewards}

The defender desires to maintain a high quality of service by controlling the voltages $V_2$  and $V_3$ near the desired normalized voltage of 1.0.  In contrast, the attacker wishes to damage equipment at node 2 by forcing $V_2$ beyond normal operating limits.  Both the defender and attacker manipulate their controls in an attempt to maximize their own average reward, expressed through the following reward functions
\begin{equation}
\label{eq:defenderReward}
R_D=-\left (\frac{V_2-1}{\epsilon}\right)^2  -\left (\frac{V_3-1}{\epsilon}\right)^2,
\end{equation}
\begin{equation}
\label{eq:attackerReward}
R_A= \Theta (V_2-(1+\epsilon)) + \Theta ((1-\epsilon)-V_2).
\end{equation}
Here, $\epsilon$ represents the halfwidth of the nominally good range of normalized voltage.  For most distribution systems under consideration, $\epsilon \sim 0.05$.  $\Theta(\cdot)$ is the step function.

\subsubsection{Players' Memory Summary Metrics}
\label{sssec:memory}

The defender and attacker use memory of the system evolution in an attempt to estimate part of the state that is not directly observable.  In principle, player memories should be constructed based on specific application domain knowledge or interviews with such experts.  However, in this initial work, we simply propose a memory summary metric for each player that potentially provides him with additional, yet imperfect, system information.  We define the defender memory summary metric to be
\begin{equation}
\label{eq:defenderMemory}
\mathcal{M}_{D,t}= \frac{1}{m+1}\sum_{n=t-m}^t\textrm{sign}(V_{1,n}-V_{1,n-1})\;\textrm{sign}(V_{3,n}-V_{3,n-1})
\end{equation}
If the attacker has very limited $q_3$ capability, both $p_3$ and $q_3$ are relatively  constant, and {\it changes} in $V_3$ should follow {\it changes} in $V_1$, which is directly controlled by the defender. If all $V_3$ changes are as expected, then $\mathcal{M}_D=1$.  The correlation between $V_1$ and $V_3$ changes can be broken by an attacker with high $q_3$ capability because large changes in $q_3$ can make $V_1$ and $V_3$ move in opposite directions. If attacker actions always cause $V_1$ and $V_3$ to move in opposite directions, then $\mathcal{M}_D=-1$. This correlation can also be broken by variability in the (unobserved) $p_2$ and $q_2$. The attacker could use this ($p_2,q_2$)  variability, which is unobserved by the attacker, to mask his actions at node 3. Such masking is more important in a setting where the defender is uncertain of the presence of the attacker, which we will address in future work.

As with the defender memory summary metric, the intent of $\mathcal{M}_A$ is to estimate some unobserved part of the state.  Perhaps the most important unobserved state variable for the attacker is $V_1$ which reveals the vulnerability of the defender and would be extremely valuable information for the attacker.  If the attacker knows the rules that the defender must follow, i.e., Equation~(\ref{eq:defendermovespace}), he can use his observations to infer $V_1$.  One mathematical construct that provides this inference is
\begin{equation}
\label{eq:attackerMemory}
\mathcal{M}_{A,t}=\sum_{n=t-m}^t\textrm{sign}\left (\textrm{floor} \left ( \frac{\Delta V_{3,n}-\Delta q_{3,n} x_2/V_0}{\delta v} \right ) \right ).
\end{equation}
If the attacker increases $q_3$ by $\Delta q_{3,t}=q_{3,t}-q_{3,t-1}$, he would expect a proportional increase in $V_3$ by $\Delta V_{3,t}=V_{3,t}-V_{3,t-1}\sim \Delta q_3 x_2/V_0$.  If $V_3$ changes according to this reasoning, then the argument in $\mathcal{M}_A$ is zero.  However, if the defender adjusts $V_1$ at the same time step, the change in $V_3$ would be modified.  If $\Delta V_{3,t}$ is greater or lower than the value expected by the attacker by $\Delta V/N$, the argument in $\mathcal{M}_A$ is +1 or -1, respectively.  The sum then keeps track of the net change in $V_1$ over the previous $m$ time steps.  Note also that the stochastic load $(p_2,q_2)$ will also cause changes in $V_3$ and, if large enough, it can effectively mask the defender behavior from the attacker. 
\subsection{Iterated Semi Network-Form Game Model}
\label{ssec:gridAsGame}

\begin{figure}[h]
\begin{center}
\includegraphics[trim=0cm 3cm 0cm 0cm, clip=true, totalheight=0.5\textheight]{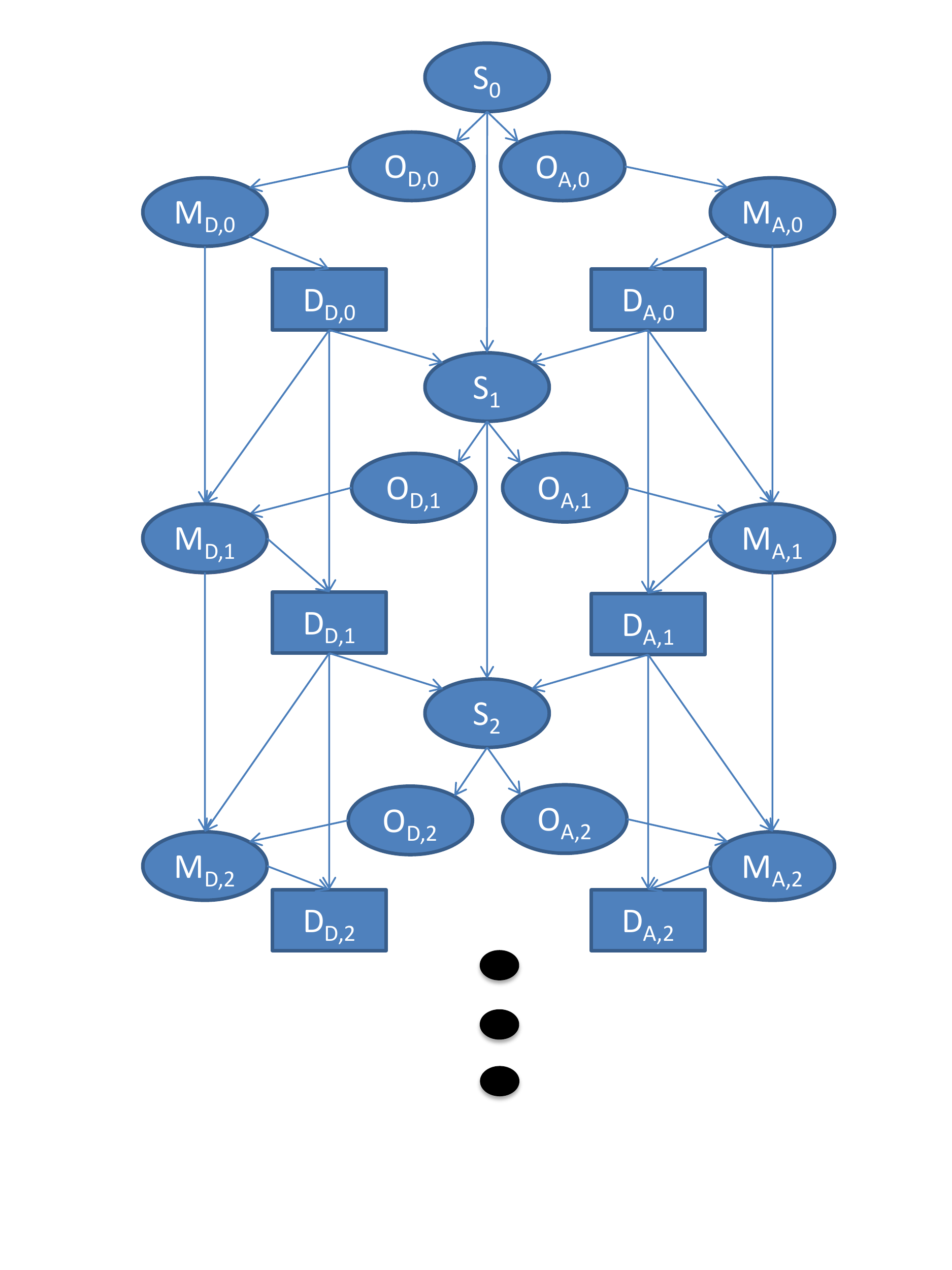}
\end{center}
\caption{The iterated semi net-form game graph of the cyber security of a smart power network scenario.  The graph shows 2 time steps explicitly.  In our experiments we choose the number of time steps $T=100$.  We use subscripts $D$ and $A$ to denote node association with the defender and attacker, respectively, and the numeric subscript to denote the time step.  The system state $S$ represents the current state of the power grid network.  The players make partial observations $O$ of the system and use them to update their memories $M$.  The memories are used to pick their action $D$.}
\label{fig:PowergridBnet}
\end{figure}

We model the scenario described in Section~\ref{ssec:gridEquations} as an iterated semi net-form game set in the graph shown in Figure~\ref{fig:PowergridBnet}.  The figure shows the net for 2 time steps with the numeric subscript on each variable denoting the time step to which it belongs.  The system state $S=[P_2, Q_2, P_1, V_1, V_2, V_3]$ is a vector that represents the current state of the power grid network.  The vector comprises of key system variables with their relationships defined in Equations~(\ref{eq:powerEquation1}) and~(\ref{eq:powerEquation2}).  The observation nodes $O_D=[V_1, V_2, V_3, P_1, Q_1]$ and $O_A=[V_2, V_3, p_3, q_3]$ are vectors representing the part of the system state that is observed by the defender and attacker, respectively.  We compute these observation nodes by taking the system state $S$, and passing through unchanged only the variables that the player observes.  Each player's observation is incorporated into a memory node ($M_D$ and $M_A$ for the defender and attacker, respectively) that summarizes information from the player's past and present.  The memory nodes\footnote{To be technically correct, we must also include the variables carried by the memory nodes $M_D$ and $M_A$ for the sole purpose of calculating $\mathcal{M}_D$ and $\mathcal{M}_A$, respectively.  However, for simplicity, we are not showing these variables explicitly.} are given by $M_{D,t}=[O_D, \mathcal{M}_{D,t}, D_{D,t-1}]$ and $M_{A,t}=[O_{A,t}, \mathcal{M}_{A,t}, D_{A,t-1}]$.  Now, the defender uses his memory $M_D$ to set the decision node $D_D$, which adjusts the setting of the voltage-tap transformer (up to one increment in either direction) and sets the voltage $V_1$.  On the other hand, the attacker uses his memory $M_A$ to set the decision node $D_A$, which sets $q_3$.  Finally, the decisions of the players are propagated to the following time step to evolve the system state.  In our experiments we repeat this process for $T=100$ time steps.

\subsection{Computing the Solution Concept}
\label{ssec:gridSolution}

We compute the level-K policies of the players following the level-K reinforcement learning solution concept described in Section~\ref{ssec:lkrl}.  First, we form the base of the level-K hierarchy by defining level 0 policies for the defender and attacker.  Then, we describe the details of how we apply reinforcement learning to bootstrap up to levels $K>0$. A level 0 policy represents a prior on the player's policy, i.e., it defines how a non-strategic player would play.  In this work, we handcrafted level 0 policies based on expert knowledge of the domain.  In future work, we would like to devise an automated and somewhat ``principled" way of setting the level 0 policies.

\subsubsection{Level 0 Policies}
\label{sssec:l0defender}

Often, level 0 players are assumed to choose their moves randomly from their move spaces $D_{D,t}$ and $D_{A,t}$.  However, we do not believe this to be a good assumption, especially for SCADA operators.  These operators have training which influences how they control the system when no attacker is present, i.e., the ``normal'' state.  In contrast, a random-move assumption may be a reasonable model for a level 0 attacker that has more knowledge of cyber intrusion than of manipulation of the electrical grid.  However, we assume that our level 0 attacker also has some knowledge of the electrical grid.

If there is no attacker present on the SCADA system, the defender can maximize his reward by adjusting $V_1$ to move the average of $V_2$ and $V_3$ closer to 1.0 without any concern for what may happen in the future.  We take this myopic behavior as representative of the level 0 defender, i.e.,
\begin{equation}
\pi_{D}(V_{2,t},V_{3,t})=   {\arg\min}_{D_{D,t}}   \frac{(V_{2,t}+V_{3,t})}{2}-1
\end{equation}

For the level 0 attacker, we adopt a \textit{drift-and-strike} policy which requires some knowledge of the physical circuit and power flow equations. We propose that the attacker ``drifts'' in one direction  by steadily increasing (or decreasing) $q_3$ by one increment at each time step.  The level 0 attacker decides the direction of the drift based on $V_2$, i.e., the attacker drifts to larger $q_3$ if $V_2<1$.  The choice of $V_2$ to decide the direction of the drift is somewhat arbitary.  However, this is simply assumed level 0 attacker behavior.  The drift in $q_3$ causes a drift in $Q_1$ and, without any compensating move by the defender, a drift in $V_2$.  However, a level 0 defender compensates by drifting $V_1$ in the opposite sense as $V_2$ in order to keep  the average of $V_2$ and $V_3$ close to 1.0. The level 0 attacker continues this slow drift until, based on his knowledge of the power flow equations and the physical circuit, he detects that a sudden large change in $q_3$ in the opposite direction of the drift would push $V_2$ outside the range $[1-\varepsilon,1+\epsilon]$.  If the deviation of $V_2$ is large enough, it will take the defender a number of time steps to bring $V_2$ back in range, and the attacker accumulates reward during this recovery time. More formally this level 0 attacker policy can be expressed as
\begin{algorithm}{Level0Attacker}{}
 V^* = \max_{q \in D_{A,t}} |V_2-1|;\\
\begin{IF}{V^* > \theta_A}
  \RETURN \arg\max_{q \in D_{A,t}} |V_2-1|;
\end{IF}\\
\begin{IF}{V_2 < 1}
  \RETURN q_{3,t-1}+1;
\end{IF}\\
\RETURN q_{3,t-1}-1;
\end{algorithm}
Here, $\theta_A$ is the threshold parameter that triggers the strike. Throughout this work, we have used $\theta_A=0.07> \epsilon $ to indicate when an attacker strike will accumulate reward.

\subsubsection{Reinforcement Learning Details}
\label{sssec:rldetails}

The training environment of a level-K player consists of all nodes that he does not control, including all chance nodes and the decision nodes of other players, which are assumed to be playing with a level $K-1$ policy.  This leaves us with a standard single-agent reinforcement learning problem, where given an observation, the player must choose an action to maximize some notion of cumulative reward.  We follow loosely the SARSA reinforcement learning setup in \cite{SuttonBook}.  First, we choose the optimization objective to be his expected sum of discounted single-step rewards (given by Equations~\ref{eq:defenderReward} and \ref{eq:attackerReward}).  To reduce the output space of the player, we impose an $\varepsilon$-greedy parameterization on the player's policy space.  That is, the player plays what he thinks is the ``best" action with probability $1-\varepsilon$, and plays uniformly randomly over all his actions with probability $\varepsilon$.  Playing all possible actions with nonzero probability ensures sufficient exploration of the environment space for learning.  At the core of the SARSA algorithm is to learn the ``Q-function", which is a mapping from observations and actions to expected sum of discounted rewards (also known as ``Q-values").  Given an observation of the system, the Q-function gives the long-term reward for playing a certain action.  To maximize the reward gathered, the player simply plays the action with the highest Q-value at each step.  

To learn the Q-function, we apply the one-step SARSA on-policy algorithm in~\cite{SuttonBook}.\footnote{Singh et al.~\cite{Singh94} describes the characteristics of SARSA when used in partially observable situations.  SARSA will converge to a reasonable policy as long as the observed variables are reasonably Markov.}  However, since the players' input spaces are continuous variables, we cannot use a table to store the learned Q-values.  For this reason, we approximate the Q-function using a neural-network~\cite{BusoniuBook,Rummery94}.  Neural networks are a common choice because of its advantages as a universal function approximator and being a compact representation of the policy.  

To improve stability and performance, we make the following popular modifications to the algorithm:  First, we run the algorithm in semi-batch mode, where training updates are gathered and updated at the end of the episode rather than following each time step.  Second, we promote initial exploration using optimistic starts (high initial Q-values) and by scheduling the exploration parameter $\varepsilon$ to a high rate of exploration at first, then slowly decreasing it as the training progresses.

\subsection{Results and Discussion}

\begin{figure}[h]
\begin{center}
\includegraphics[totalheight=0.9\textheight]{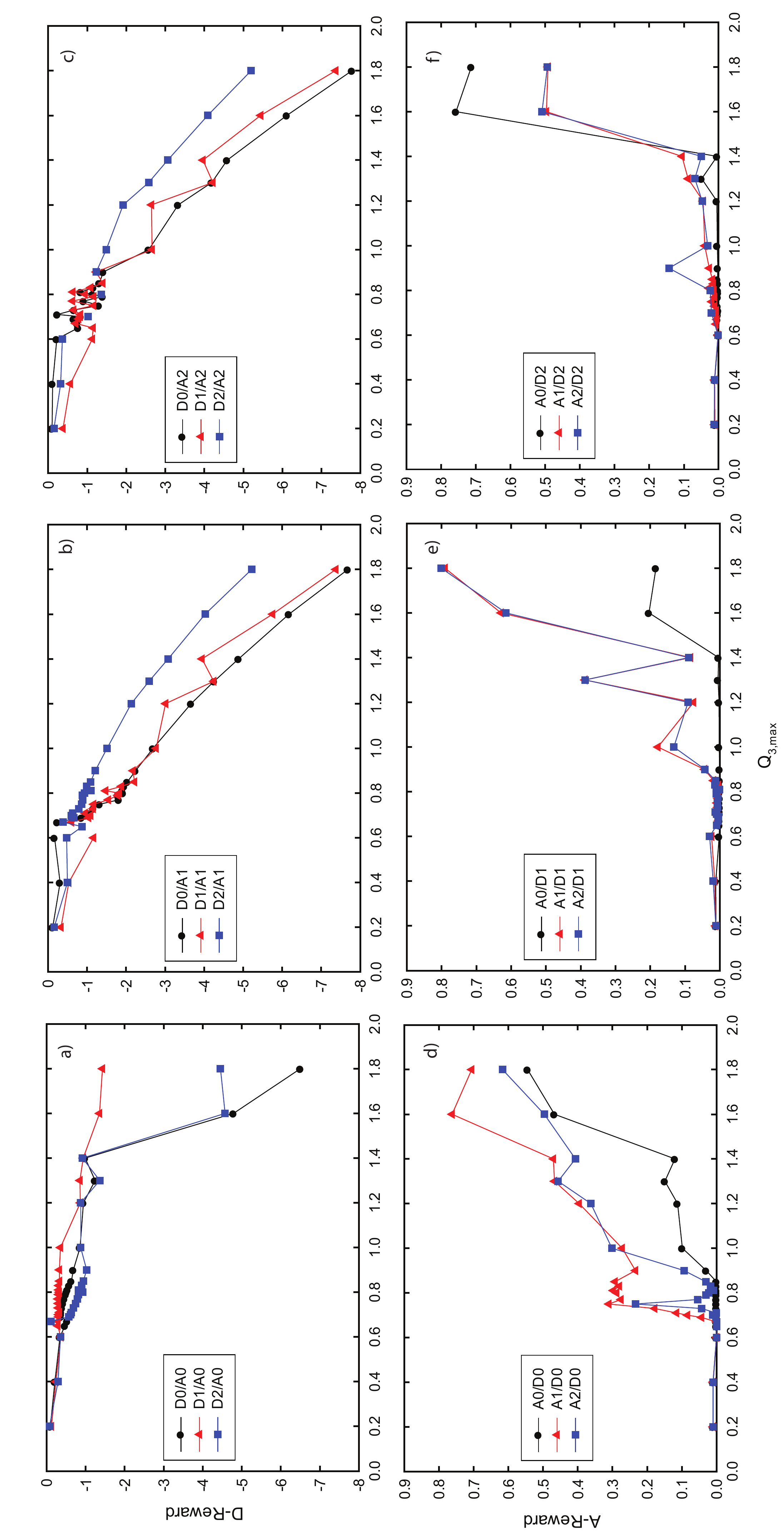}
\end{center}
\caption{Average reward per step averaged over 50 episodes as a function of $q_{3,max}$ for all pairings of the defender (D) and attacker (A) through level 2. (a) Reward of D0, D1, and D2 when matched against A0.  (b) Same as (a) but for A1.  (c) Same as (a) and (b) but for A2.  (d) Reward of A0, A1, and A2 when matched against D0. (e) Same as (d) but for D1.  (f) Same as (d) and e) but for D2.  In general, we observe that as $q_{3,max}$ increases, the defender's average reward decreases and the attacker's average reward increases.}\label{fig:performance}
\end{figure}

\begin{figure}[h]
\begin{center}
\includegraphics[width=1.0\linewidth]{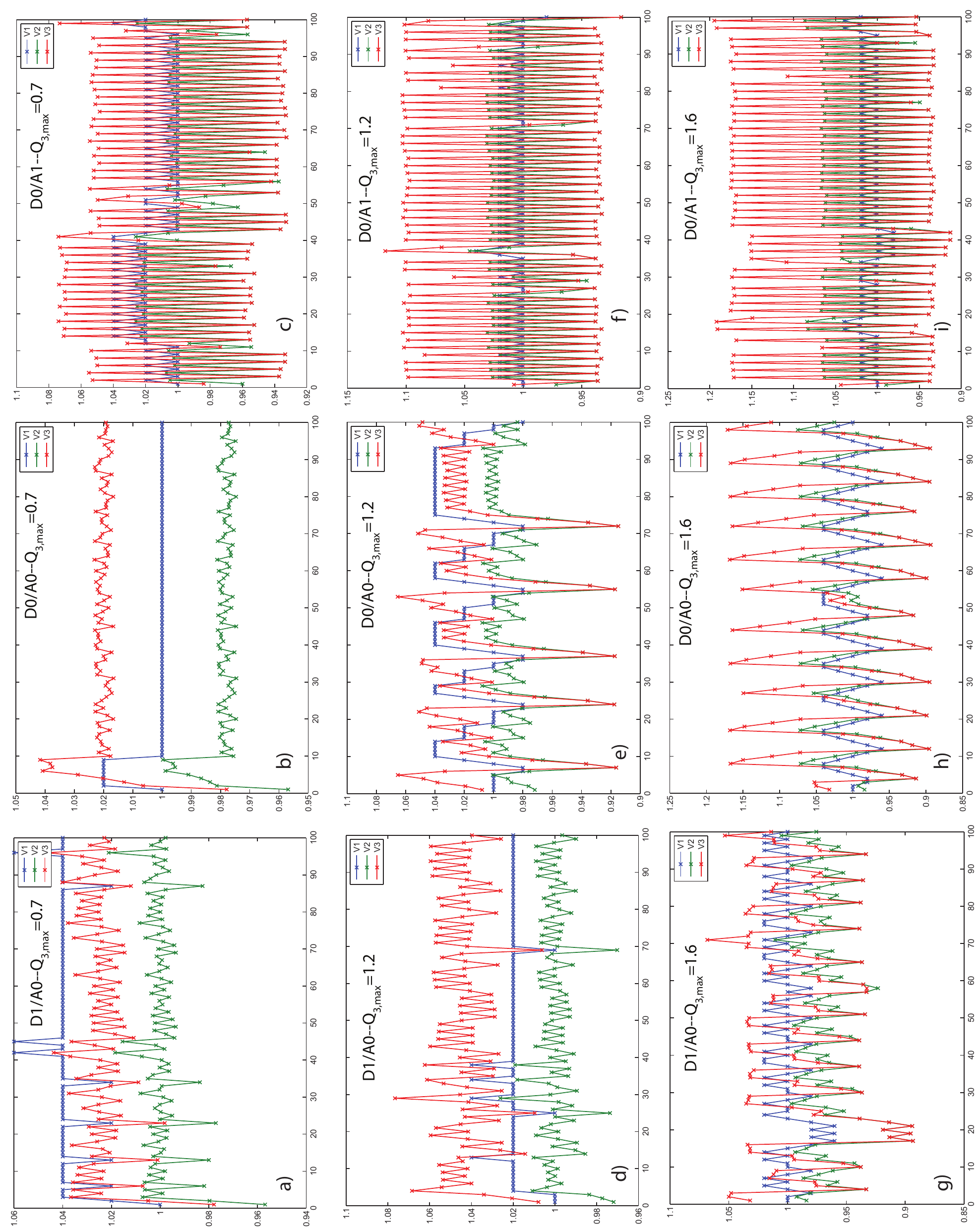}
\end{center}
\caption{Simulations of system voltages for level 0 and level 1 that show the evolution in level 1 attacker (A1) and level 1 defender (D1) policies after reinforcement learning training session against their level 0 counterparts D0 and A0.  (a) D1 versus A0, (b) D0 versus A0, and (c) D0 versus A1 for $q_{3,max}=0.7$.  (d) D1 versus A0, (e) D0 versus A0, and (f) D0 versus A1 for $q_{3,max}=1.2$. (g) D1 versus A0, (h) D0 versus A0, and (i) D0 versus A1 for $q_{3,max}=1.6$. In the center column (D0 versus A0), the attacker becomes increasingly capable of scoring against the defender as $q_{3,max}$ is increased.  In the left column (D1 versus A0), the defender is successful at avoiding attacks by not chasing small immediate rewards from voltage centering.  In the right column (D0 versus A1), the attacker successfully leverages the level 0 defender's move to help him score.}
\label{fig:level-1-behavior}
\end{figure}

\begin{figure}[h]
\begin{center}
\includegraphics[width=1.0\linewidth]{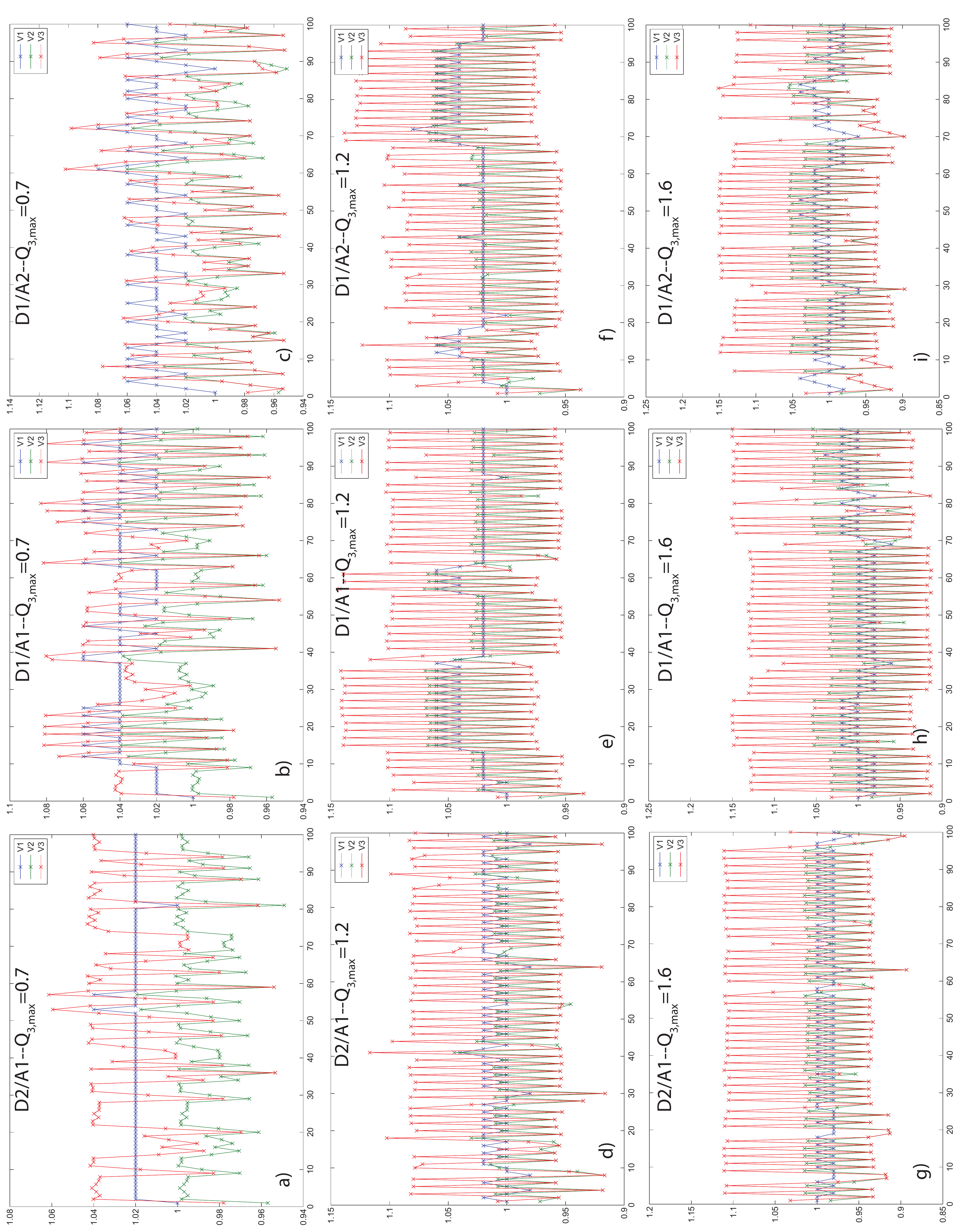}
\end{center}
\caption{Simulations of system voltages for level 1 and level 2 that show the evolution in level 2 attacker (A2) and level 2 defender (D2) policies after reinforcement learning training session against their level 1 counterparts D1 and A1.  (a) D2 versus A1, (b) D1 versus A1, and (c) D1 versus A2 for $q_{3,max}=0.7$.  (a) D2 versus A1, (b) D1 versus A1, and (c) D1 versus A2 for $q_{3,max}=1.2$. (g) D2 versus A1, (h) D1 versus A1, and (i) D1 versus A2 for $q_{3,max}=1.6$.}\label{fig:level-2-behavior}
\end{figure}

Level-K reinforcement learning was performed for all sequential combinations of attacker and defender pairings, i.e., D1/A0, D2/A1, A1/D0, and A2/D1.  Here, we refer to a level $K$ player using a shorthand where the letter indicates attacker or defender and the number indicates the player's level.  The pairing of two players is indicated by a ``/''.  The training was performed for $q_{3,max}$ in the range 0.2 to 1.8.  Subsequent to training, simulations were run to assess the performance of the different player levels.  The player's average reward per step for the different pairs is shown in Figure~\ref{fig:performance} as a function of $q_{3,max}$.  Figure~\ref{fig:level-1-behavior} shows snapshots of the players' behavior for the pairings D0/A0, D1/A0, and D0/A1 for $q_{3,max}=$0.7, 1.2, and 1.6.  Figure~\ref{fig:level-2-behavior} shows the same results but for one level higher, i.e., D1/A1, D2/A1, and D1/A2.

\paragraph{D0/A0}Figures~\ref{fig:level-1-behavior}(b), (e), and (h) show the interaction between the two level 0 policies, and Figures~\ref{fig:performance}(a) and (d) show the average player performance.  These initial simulations set the stage for interpreting the subsequent reinforcement learning.  For $q_{3,max}<0.8$, the black circles in Figure~\ref{fig:performance}(d) show that A0 is unable to push $V_2$ outside of the range $[1-\epsilon, 1+\epsilon]$.  The explanation is found in Figure~\ref{fig:level-1-behavior}(b).  With $V_2<1$ and say $q_{3,max}=0.7$, A0's drift will have saturated at $q_3=q_{3,max}=0.7$.  However, with $\theta_\mathcal{A}=0.07$, A0 will not strike by changing $q_3=-q_{3,max}=-0.7$ unless he projects such a strike could drive $V_2$ below 0.93.  A0's limited $q_3$-strike capability is not enough overcome the threshold and the system becomes locked in a quasi-steady state.  In the midrange of A0's capability ($0.8\leq q_3 \leq 1.4$), the drift-and-strike A0 policy is effective (Figure~\ref{fig:level-1-behavior}(e)).  However, A0 is only successful for strikes that force $V_2<0.95$.  In addition, there are periods of time when $V_2\sim 1.0$ and A0 is unable to decide on a drift direction. However, these become fewer (and A0's average reward grows) as $q_{3,max}$ approaches 1.4 (Figure~\ref{fig:performance}(d)).  For $q_{3,max}\geq 1.6$, A0 is able to successfully strike for $V_2<0.93$ and $V_2>1.07$, and A0 drives the system into a nearly periodic oscillation (Figure~\ref{fig:level-1-behavior}(h)) with a correspondingly large increase in A0's discounted average reward (Figure~\ref{fig:performance}(d)).  The reduction in D0's performance closely mirrors the increase in A0's performance as $q_3$ increases.  However, {\it it is important to note that D0 enables much of A0's success by changing $V_1$ to chase the $V_2$ and $V_3$.}  The adjustments in $V_1$ made by D0 in Figures~\ref{fig:level-1-behavior}(b), (e), and (h) bring the system closer to the voltage limits just as A0 gains a large strike capability.

\paragraph{D1 Training Versus A0}The red triangles in Figure~\ref{fig:performance}(a) and the black circles in Figure~\ref{fig:performance}(e) show dramatic improvement in the performance of D1 over D0 when faced with A0.  In the middle range of A0's capability ($0.8\leq q_{3,max} \leq 1.4$), Figure~\ref{fig:level-1-behavior}(d) shows that D1 stops changing $V_1$ to chase the immediate reward sought by D0.  Instead, D1 maintains a constant $V_1=1.02$ keeping $V_2\sim 1.0$ and A0 uncertain about which direction to drift.  By keeping $V_1>1.0$, D1 also corrects the error of D0 whose lower values of $V_1$ helped A0 push $V_2$ and $V_3$ below $1-\epsilon$.  With $V_1=1.02$, the average of $V_2$ and $V_3$ are significantly higher than 1.0, but D1 accepts the immediate decrement in average reward to avoid a much bigger decrement he would suffer from an A0 strike.  The effect of this new strategy is also reflected in the poor A0 performance as seen from the black circles in Figure~\ref{fig:performance}(e).  The behavior of D1 for $q_{3,max}\geq 1.6$ in Figure~\ref{fig:level-1-behavior}(g) becomes complex.  However, it appears that D1 has again limited the amount that he chases $V_2$ and $V_3$.  In fact, D1 moves $V_1$ in a way that decreases his immediate reward, but this strategy appears to anticipate A0's moves and effectively cuts off and reverses A0 in the middle of his drift sequence.  We note that this behavior of the defender makes sense because he knows that the attacker is there waiting to strike.  In real life, a grid operator may not realize that a cyber attack is even taking place.  To capture this phenomenon motivates follow-on work in uncertainty modeling of the attacker's existence.

\paragraph{A1 Training Versus D0}A cursory inspection of Figures~\ref{fig:level-1-behavior}(c), (f), and (i) might lead one to believe that the A1 training has resulted in A1 simply oscillating $q_3$ back and forth from $+q_{3,max}$ to $-q_{3,max}$.  However, the training has resulted in rather subtle behavior, which is most easily seen in Figure~\ref{fig:level-1-behavior}(c).  The largest change A1 (with $q_{3,max}=0.7$) can independently make in $V_2$ is $\sim 0.04$.  However, A1 gains an extra 0.02 of voltage change by leveraging (or perhaps convincing) D1 to create oscillations of $V_1$ in-phase with his own moves.  For this strategy to be effective in pushing $V_2$ below $1-\epsilon$, the $V_1$ oscillations have to take place between 1.0 and 1.02, or lower.  When the synchronization of the $V_1$ and A1 oscillations are disturbed such as at around step 50 in Figure~\ref{fig:level-1-behavior}(c), A1 modifies his move in the short term to delay the move by D0 and  re-establish the synchronization.  A1 also appears to have a strategy for ``correcting'' D0's behavior if the oscillations take place between levels $V_1$ that are too high.  Near step 40 in Figure~\ref{fig:level-1-behavior}, A1 once again delays his move convincing D0 to make two consecutive downward moves of $V_1$ to re-establish the ``correct'' D0 oscillation level.  Similar behavior is observed out to $q_{3,max}=1.4$.  At $q_{3,max}=1.6$, A1 has enough capability that he can leverage in-phase D0 oscillations to exceed both the $V_2$ lower and upper voltage limits.  This improved performance is reflected in the dramatic increase in A1's average reward (A1/D0; see red triangles in Figure~\ref{fig:performance}(d)).

\paragraph{D1/A1}In the hierarchy of level-K reinforcement learning, D1/A1 is similar to D0/A0 in that they do not train against one another, but this match up sets the stage for interpreting the level-2 trainings.  Figures~\ref{fig:level-1-behavior}(a), (d), and (g) show that the D1/A0 training results in a D1 that does not chase $V_2$ and $V_3$, keeps $V_2$ near 1.0, and accepts a lower current reward to avoid large A0 strikes.  In Figures~\ref{fig:level-2-behavior}(b), (e), and (h), D1 continues to avoid responding to the oscillatory behavior of A1, $V_2$ generally does not cross beyond the acceptable voltage limits.  However, $V_3$ is allowed to deviate significantly beyond the bounds.  The result is that D1's average reward versus A1 does not show much if any improvement over D0's versus A1 (red triangles and black circles, respectively, in Figure~\ref{fig:performance}(b)).  However, D1 is quite effective and reducing the performance of A1 (Figures~\ref{fig:performance}(e) red triangles) relative to the performance of A1 in D0/A1, at least for the intermediate values of $q_{3,max}$ (Figure~\ref{fig:performance}(d) red triangles).  The results for A1 are clearer.  Figures~\ref{fig:level-2-behavior}(b), (e), and (h) show the oscillatory behavior of A1 while Figures~\ref{fig:performance}(a), (b), (d), and (e) show that the switch from A0 to A1 when facing D1 improves the attacker's performance while degrading the performance of D1.

\paragraph{D2 Training Versus A1}The results of this training start out similar to the training for D1.  Figure~\ref{fig:level-2-behavior}(a) shows that, at $q_{3,max}=0.7$, D2 performs better if he does not make many changes of $V_1$ thereby denying A1 the opportunity to leverage his moves to amplify the swings of $V_2$.  For the higher values of $q_{3,max}$ in Figures~\ref{fig:level-2-behavior}(d) and (g), D2 learns to anticipate the move pattern of A1 and moves in an oscillatory fashion, but one that is {\it out of phase} with the moves of A1.  Instead of amplifying the swings of $V_2$, D2's moves attenuate these swings.  This new behavior results in across-the-board improvement in D2's average discounted reward over D1 (blue squares versus red triangles in Figure~\ref{fig:performance}(b) and a significant reduction in A1 performance (red triangles in Figure~\ref{fig:performance}(e) versus Figure~\ref{fig:performance}(f)).

\paragraph{A2 Training Versus D1}A2 shows no perceptible increase in performance over A1 when matched against D1 (blue squares versus red triangles in Figure~\ref{fig:performance}(e)).  The same general observation can be made for A2 and A1 when matched against any of D0, D1, or D2.  Figures.~\ref{fig:performance}(b) and (c) show that the defenders perform nearly the same against A1 or A2, and Figures~\ref{fig:performance}(e) and (f) show no significant change in attacker performance when switching from A1 to A2.  This may indicate that policies embodied in A2 (or A1) may be approaching a fixed point in performance.

\paragraph{D2/A2} The similarities in the performance of A1 and A2 make the analysis of this interaction nearly the same as that of D2/A1.

\acresetall
\section{Conclusions and Future Work}
\label{sec:conclusions}
%Conclusions and future work, including:
%Uncertainty of attacker's existence.  Setting L0 distributions.  Online adaptation of player policies.  Porting to libnfg.  Getting real data.
%
%Check libnfg.next.steps.11.7.11.txt.
%
%1) Conclusions for this paper
%
%2) Future modeling changes
%	a) uncertainty about existence, level, utility function, etc.
%	b) unawareness
%	c) adaptive strategies, setting L0 distributions by Laplace
%	d) interacted strategies
%	
%3) Estimation and validation
%	a) gathering data via crowd sourcing
%	b) comparing against baseline models, i.e. agent-based or equilibrium-based.
%
%4) Porting to LibNFG
%	
%5) Other projects
%	a) Erik's multi-fi project
%	b) general cyber-security project
%	c) Yildiray's nextgen cyber-security project
%	d) Juan and JB's ACBM project
%	e) Ritchie's runway incursion project

In this paper, we introduced a strategic, computationally-tractable, experimentally-motivated model for predicting human behavior in novel and complex time-extended scenarios.  This model consists of an iterated semi net-form game combined with a level-K RL solution concept.  We applied this model to predict behavior on a cyber battle on a smart power grid. As discussed in the results section, the predictions of this model are promising in that they match expectations for how a ``real world" cyber battle would unfold.  

We can vary parameters of the model that both concern the kind of cyber battle taking place (e.g., degree of compromise) and that describe the players (e.g., level 0 distributions, their level $K$).  We can also vary the control algorithm. We can then evaluate the expected ``social welfare" (i.e., the happiness metric of the system designer) for all such variations. In this way our framework can be used to increase our understanding of existing and proposed control algorithms to evaluate their robustness under different cyber attack scenarios and/or model mis-specification. In the near future, with additional advances in our computational algorithms, we hope to be able to solve the model in real-time as well. This raises the possibility of using our framework to do real-time control rather than choose among some small set of proposed control algorithms, i.e., to dynamically predict the attacker's policy and respond optimally as the cyber battle unfolds.
% \btc{We should say something exciting here about the opportunities this modeling approach opens up before we start talking about our deficits}
% JWB: I took a hack at this above.

Despite the significant modeling advances presented here, there are several important ways in which the realism of this paper's model can be improved. Some of these improvements have already been formalized, but they were left out of this document for the purposes of space and clarity. For example, the iterated semi net-form game framework easily models the situation where players have uncertainty about the environment they are facing. This includes uncertainty about the utility functions and the rationality (or levels) of the other players. This naturally corresponds to the Bayesian games setting within the extensive form games formalism. This also includes uncertainty about whether or not the other players exist.  In fact, the semi net-form game formalism is unique in that it can even be extended to handle ``unawareness" -- a situation where a player does not know of the possibility of some aspect of the game. 
%Unawareness is different from uncertainty because an unaware player has a fundamentally flawed (not merely inaccurate) model of the physical reality of the game in which he is involved. It is as if the unaware player has a prior that assigns zero probability to the events of which he is unaware. So the unaware player's model is flawed because no amount of Bayesian updating will make his model accurate.  
For example, it would be unawareness, rather than uncertainty, if the defender did not know of the possibility that an attacker could take control of a portion of the smart power grid.  
%Even if the defender observes an extended period of abnormal power grid behavior, he still will not suspect that the observed behavior is the result of a cyber attack if he is unaware that such a cyber attack can occur. Unawareness is a major stumbling block of conventional game theoretic approaches in part because it forces a disequilibrium.  
These types of uncertainty and unawareness will be presented and explored in future work.
%\btc{I think a bit longer explanation of why this distinction is important would be good for the CS people}
% JWB: changed and changed back after talks with David and Ritchie.

Another important modeling advance under development is related to the ability of players to adapt their policies as they interact with their opponents and make observations of their opponents' actual behavior. The level-K RL solution concept is particularly well-suited to relatively short-term interactions, like the cyber battle analyzed above. However, as interactions draw out over a longer time-frame, we would expect the players to incorporate their opponent's actual behavior into their level-K model of their opponent. One possibility for achieving this type of adaptation is based on a player using a Bayesian variant of fictitious play to set the level 0 distribution of their opponent. In other words, we use the past behavior to update the level 0 distribution of the opponent. 
% JWB: I hesitate to get into the reasons why this adaptive RL would be undesirable and/or why a PGT interacted approach would be best...

This discussion raises an important question about what happens when the strategic situation is not novel and/or the players have previously interacted. Is the level-K RL model developed here still appropriate? The answer is probably no. In such an \emph{interacted} environment, we should expect the players to have fairly accurate beliefs about each other. Furthermore, these accurate beliefs should lead to well-coordinated play.  For example, in the power grid this would mean that the attacker and defender have beliefs that correspond to what the other is actually doing rather than corresponding to some independent model of the other's behavior.  In the very least, we should not expect the players to be systematically wrong about each other as they are in the level-K model.  Rather, in this interacted environment, player behavior should be somewhere between the completely \emph{non-interacted} level-K models and a full-on equilibrium, such as Nash equilibrium or quantal response equilibrium.  The analysis of interacted, one-shot games found in Bono and Wolpert \cite{Bono11,Wolpert08} should provide a good starting point for developing a model of an interacted, time-extended game.

Perhaps the most important next step for this work is the process of estimating and validating our model using real data on human behavior.  We specifically need data to estimate the parameters of the utility functions and the level $K$ of the players as well as any parameters of their level 0 strategies.  After fitting our model to data, we will validate our model against alternative models. The difficult part about choosing alternative models with which to compare our model is that extensive-form games and equilibrium concepts are computationally intractable in the types of domains for which our model is designed. Therefore, feasible alternative models will likely be limited to simplified versions of the corresponding extensive-form game and agent-based simulations of our iterated semi net-form game.

For the smart grid cyber battle analyzed in this paper, there are several options for gathering data. One is to conduct conventional game-theoretic experiments with human subjects in a laboratory setting. Unfortunately, estimating our model, especially with the modeling advances discussed above, will require more data than is practical to collect via such conventional experimental methods which involve actual power grid operators in realistic settings. An alternative method for collecting the large amount of data required is via ``crowd-sourcing". In other words, it should be possible to deploy an internet-application version of our smart grid cyber battle to be played by a mixture of undergraduates, researchers, and power engineers. The data from these experiments would then be used to estimate and validate our model.

The methodologies presented here, and the proposed future extensions, also apply to many other scenarios. Among these are several projects related to cyber security as well as the Federal Aviation Administration's NextGen plan for modernizing the National Airspace System.  To encompass this range of applications, we are developing libNFG as a code base for implementing and exploring NFGs~\cite{Lee11}. The development of this library is ongoing, and modeling advances, like those mentioned above, will be implemented as they become an accepted part of the modeling framework. The libNFG library will ultimately be shared publicly and will enable users to fully customize their own iterated semi net-form game model and choose from a range of available solution concepts and computational approaches.
% \btc{This seems like a weird tone to end on. I think rephrasing it to be more of the tone: ``The methodologies presented here, and the proposed future extensions, also apply to many other scenarios. To encompass the range of applications, we are developing libNFG as a code base for implementing and exploring NFGs}
% JWB: incorporated above

\begin{acknowledgement}
This research was supported by the NASA Aviation Safety Program SSAT project, and the Los Alamos National Laboratory LDRD project Optimization and Control Theory for Smart Grid.
\end{acknowledgement}

%%%%%%%%%%%%%%%%%%%%%%%% referenc.tex %%%%%%%%%%%%%%%%%%%%%%%%%%%%%%
% sample references
% %
% Use this file as a template for your own input.
%
%%%%%%%%%%%%%%%%%%%%%%%% Springer-Verlag %%%%%%%%%%%%%%%%%%%%%%%%%%
%
% BibTeX users please use
% \bibliographystyle{}
% \bibliography{}
% Use the following syntax and markup for your references if
% the subject of your book is from the field
% "Mathematics, Physics, Statistics, Computer Science"
%
% Contribution
%\bibitem{science-contrib} Broy, M.: Software engineering --- from auxiliary to %key technologies. In: Broy, M., Dener, E. (eds.) Software Pioneers, pp. 10-13. %Springer, Heidelberg (2002)
%
% Online Document
%\bibitem{science-online} Dod, J.: Effective substances. In: The Dictionary of %Substances and Their Effects. Royal Society of Chemistry (1999) Available via %DIALOG. \\
%\url{http://www.rsc.org/dose/title of subordinate document. Cited 15 Jan 1999}
%
% Monograph
%\bibitem{science-mono} Geddes, K.O., Czapor, S.R., Labahn, G.: Algorithms for %Computer Algebra. Kluwer, Boston (1992)
%
% Journal article
%\bibitem{science-journal} Hamburger, C.: Quasimonotonicity, regularity and %duality for nonlinear systems of partial differential equations. Ann. Mat. Pura. %Appl. \textbf{169}, 321--354 (1995)
%
% Journal article by DOI
%\bibitem{science-DOI} Slifka, M.K., Whitton, J.L.: Clinical implications of dysregulated cytokine production. J. Mol. Med. (2000) doi: 10.1007/s001090000086
%
%\bigskip

%\bibliographystyle{unsrt}
%\bibliography{RLBib,ThesisBib,SmartGrid}

\end{document}